\documentclass[showpacs,preprintnumbers,amsmath,amssymb,12pt]{article}
\usepackage{graphicx,amsfonts,amssymb,amsthm,amsmath,graphics,epsfig,pstricks,fancyhdr,fancybox}
\usepackage{dcolumn}
\usepackage{bm}
\usepackage{bbm}
\usepackage{a4wide}
\usepackage{graphicx}
\usepackage{subfigure}
\usepackage{dsfont}
\newtheorem{thm}{Teorema}[section]

\newtheorem{prop}[thm]{Proposition}

\newcommand{\rv}{\textit{rv}}
\newcommand{\id}{\textit{id}}
\newcommand{\iid}{\textit{iid}}
\newcommand{\sd}{\textit{sd}}

\newcommand{\pdf}{\textit{pdf}}
\newcommand{\cdf}{\textit{cdf}}

\newcommand{\chf}{\textit{chf}}

\newcommand{\PqoP}{\mathbb{P}\hbox{-\emph{a.s.}}}
\newcommand{\eqd}{\stackrel{d}{=}}

\newcommand{\poiss}{\mathfrak{P}}

\newcommand{\erl}{\mathfrak{E}}

\newcommand{\bin}{\mathfrak{B}}

\newcommand{\indim}{\noindent{\bf Proof:}\hspace{0.2cm}}
\newcommand{\findim}{\hfill$\square$\vspace{0.3cm}\noindent}

\def\ito{It\={o}}

\begin{document}
\thispagestyle{empty}

\title{\Huge \textbf{Cointegrating Jumps: \\
an Application to Energy Facilities\footnote{The technique here discussed does not reflect UGC view.}}
\author{Nicola \textsc{Cufaro Petroni}\footnote{cufaro@ba.infn.it}  \\
Dipartimento di \textsl{Matematica} and \textsl{TIRES}, Universit\`a di Bari\\
\textsl{INFN} Sezione di Bari\\ \vspace{7pt}
via E. Orabona 4, 70125 Bari, Italy\\
Piergiacomo \textsc{Sabino}\footnote{piergiacomo.sabino@uniper.energy}\\
\textsl{RQPR} Quantitative Risk Modelling and Analytics \\
\textsl{Uniper} Global Commodities SE\\
\vspace{5pt}
 Holzstrasse 6, 40221 D\"usseldorf, Germany
} }

\date{}

\maketitle

\begin{abstract}
\noindent Based on the concept of self-decomposable random variables we
discuss the application of a model for a pair of dependent Poisson
processes for energy facilities. Due to the resulting structure of
the jump events, we can see the self-decomposability as a form of
cointegration among jumps. In the context of energy facilities, the
application of our approach to model power or gas dynamics and to
evaluate transportation assets seen as spread options is
straightforward. We study the applicability of our methodology first
assuming a Merton market model with two underlying assets; in a
second step we consider price dynamics driven by an exponential
mean-reverting Geometric Ornstein-Uhlenbeck  plus compound Poisson
that are commonly used in the energy field. In this specific case we
adopt a price spot dynamics  for each underlying that has the
advantage of being treatable to find no-arbitrage conditions. In
particular we can find close-form formulas for vanilla options
so that the price and the Greeks of
spread options can be calculated in close form using the Margrabe
formula \cite{Mar78} (if the strike is zero) or some well known
approximations as in Deng et al. \cite{DLZ2006}.
\end{abstract}

\section{Introduction and Motivation}
Several research studies have shown that the spot dynamics of commodity prices is subjected to mean reversion, seasonality and jumps (see for example Cartea and Figueroa \cite{CarteaFigueroa}). In addition, some methodologies have also been proposed to take dependency into account based on correlation and cointegration. However, these approaches can become too mathematically complex or non-treatable when leaving the Gaussian-{\ito} world.

In this paper we address the problem of dependency in the
$2$-dimensional case and start considering $2$-dimensional jump
diffusion processes with a $2$-dimensional compound Poisson
component. We then introduce an intuitive approach to model the
dependency of $2$-dimensional Poisson processes based on the
self-decomposability (see Cufaro
Petroni~\cite{Cufaro08}, Cufaro Petroni and Sabino~\cite{cs15},
Sato~\cite{Sato}) of the exponential random variables used for its
construction. In Cufaro Petroni and Sabino~\cite{cs15} we have seen that given two independent exponential \rv's
$Y,Z\sim\erl(\lambda)$, and a $0\,$-$1$ Bernoulli \rv\
$B(1)\sim\bin(1,1-a)$ with $a=\mathbf{P}\left\{B(1)=0\right\}$, then
also the \rv\ defined as
\begin{equation}
    X=aY+Z_a\qquad\quad Z_a=B(1)Z
\end{equation}
is an exponential $\erl(\lambda)$ resulting in a weighed sum of $Y$
and $Z$ where $0<a<1$ is the \emph{deterministic} weight of $Y$,
while $B(1)$ is the \emph{random} weight of $Z$. In other words $X$
is nothing else than the exponential $Y$ down $a$-rescaled, plus
another independent, but \emph{intermittent} with frequency $1-a$,
exponential $Z$. It is apparent on the other hand that, by
construction, $X$ and $Y$ are not independent and it is possible to
show that $a$ also represents precisely their correlation
coefficient. This result is a direct consequence of the
self-decomposability of the exponential laws.

As a matter of fact we could also produce pairs of
$a$-correlated exponentials $X\sim\erl(\lambda)$ and
$Y'\sim\erl(\mu)$ with different parameters by reformulating the
previous relation as
\begin{equation}
    X = \gamma Y' + Z_a
\end{equation}
where $\gamma = \mu a/\lambda$, so that
$0<\gamma<\frac{\mu}{\lambda}$. Considering then  $X$ and $Y'$ as
two random times with a positive
random delay $Z_a$, the mathematical concept of self-decomposability
can help describing their co-movement and can answer some common
questions arising in the financial context:
\begin{itemize}
    \item Once a financial institution defaults how long should one wait for a dependent institution to default too?
    \item  A market receives a news interpreted as a shock: how long should one wait to see the propagation of that shock onto a dependent market?
    \item If different companies are interlinked, what is the impact on insurance risk?
\end{itemize}
Questions like the ones above are covered by the special case
$\gamma > 1$. Our model is then rich enough to describe cases where
the second random time event occurs before the first
one. Similar results based on linear structure of exponential \rv\
can be found in Iyer et al. \cite{IMM2002} whose purpose was to
model a multi-component reliability system.

It is worthwhile to notice that this bivariate exponential model
implies a copula function (see
Cufaro Petroni and Sabino~\cite{cs15}) that is neither chosen
upfront nor whose parameters are estimated from market
data: here the copula function does
not define the model, but rather the opposite.

Based on the self-decomposability of the exponentional \rv's we are able to construct $2$-dimensional Poisson processes with dependent marginals (see Cufaro Petroni and Sabino \cite{cs15}). Because of the relationship among random times, the two Poisson processes can be seen linked with a form of cointegration between their jumps.

In the context of energy facilities, the application of our approach to model price dynamics and to evaluate transportation assets seen as spread options is straightforward. To this purpose, we consider the German EEX and French Powernext power markets and assume that each spot price dynamics is driven by an exponential mean-reverting Geometric Ornstein-Uhlenbeck (GOU) plus compound Poisson. In this specific case we adopt a stochastic dynamics of the spot prices that is slightly different from the one in Cartea and Figueroa \cite{CarteaFigueroa} with the advantage of being more treatable to find no-arbitrage conditions. In particular we can find close-form formulas for vanilla options hence the price and the Greeks of  spread options can be calculated in close form using the Margrabe formula \cite{Mar78} (if the strike is zero) or some well known approximations as in Deng et al. \cite{DLZ2006}, Pellegrino and Sabino \cite{PellegrinoSabino_2} and Pellegrino \cite{Pellegrino2016}. In any case our approach implies an explicit algorithm for the simulation of the dependent Poisson processes and can be used in Monte Carlo simulations.

Finally, we compare the results obtained by our approach to the ones obtained assuming that the two compound Poisson processes are independent or contain a common Poisson component.

The extension to the multi-dimensional case will be the goal of future studies as well as the extension to different dynamics other than Poisson. However, under the assumption that only two underlyings have jump component, the price and the Greeks of spread options can be obtained by the moment-matching methodology proposed in Pellegrino and Sabino \cite{PellegrinoSabino_1}.

The paper is organized as follows. Section \ref{sec:bi_poisson} summarizes the results for bi-dimensional Poisson process introduced in Cufaro Petroni and Sabino \cite{cs15}. In section \ref{sec:The Models} we consider bi-dimensional jump diffusion processes having a Geometric Brownian Motion (GBM) and GOU diffusive component. We also  apply our methodology to the $2$-factor Schwartz-Smith model \cite{SchwSchm00} with jump diffusion where we find analytical solutions for vanilla options as well. Section \ref{sec:rn_pricing} presents the risk neutral formulas for plain vanilla and spread options given the price dynamics introduced in Section \ref{sec:The Models} and given the different types of bi-dimensional Poisson components. Section \ref{sec:numerical_experiments} illustrates our approach with practical examples: we first assume a pure GBM plus jump model (Merton model) and compare the results obtained by our approach to the ones obtained assuming that the two compound Poisson processes are independent or contain a common Poisson component. In a second step we estimate the parameters of a $2$-dimensional GOU plus jumps dynamics to model the EEX and Powernext day-head prices. Finally we compare the price of an interconnection between these two locations assuming the three types of Poisson configuration mentioned above. Section \ref{sec:Conclusions} concludes the paper with an overview of future studies and possible further applications.

\section{Dependent Poisson processes}\label{sec:bi_poisson}

A law with density (\pdf) $f(x)$ and characteristic function (\chf)
$\varphi(u)$ is said to be \emph{self-decomposable} (\sd) (see
Sato~\cite{Sato} and Cufaro Petroni~\cite{Cufaro08}) if for every
$0<a<1$ we can find another law with \pdf\ $g_a(x)$ and \chf\
$\chi_a(u)$ such that
\begin{equation*}
    \varphi(u)=\varphi(au)\chi_a(u)
\end{equation*}
This definition selects an important
family of laws with many relevant properties. Remark however that,
while a \sd\ $\chi_a(u)$ can be explicitly expressed in terms of
$\varphi(u)$, its corresponding \pdf\ $g_a(x)$ can not be given in a
general, elementary form from $f(x)$. We will also say that a random
variable (\rv) $X$ is \sd\ when its law is \sd: looking at the
definition this means that for every $0<a<1$ we can always find two
\emph{independent} \rv's $Y$ (with the same law of $X$), and $Z_a$
with \pdf\ $g_a(x)$ and \chf\ $\chi_a(u)$ such that \emph{in
distribution}
\begin{equation*}
    X\eqd aY+Z_a
\end{equation*}
We can look at this, however, also from a different point of view:
to the extent that for $0<a<1$ the law of $Z_a$ is known, we can
define the \rv
\begin{equation*}
   X=aY+Z_a
\end{equation*}
which by self-decomposability will now have the same law of $Y$. It
would be easy to show that $a$ also plays the role of the
correlation coefficient between $X$ and $Y$, namely
\begin{equation*}
    r_{XY}=a
\end{equation*}

It is well known, in particular, that the exponential laws
$\erl_1(\lambda)$ with \pdf\ and \chf
\begin{equation*}
    f_1(x)=\lambda e^{-\lambda x}\mathbbm{1}_{x\ge 0} \qquad\qquad \varphi_1(u)=\frac{\lambda}{\lambda-iu}
\end{equation*}
are a typical example of \sd\ laws (see Sato \cite{Sato}).

It is possible to show now
(Cufaro Petroni and
Sabino~\cite{cs15}) that the law of $Z_a$ is a mixture of a law
$\bm\delta_0$ degenerate in $0$, and an exponential
$\erl_1(\lambda)$, namely
\begin{equation*}
    Z_a\sim a\bm\delta_0+(1-a)\erl_1(\lambda)
\end{equation*}
and this entails that $Z_a$ can be taken as the product of two \id\
\rv's: $Z\sim\erl_1(\lambda)$, and
$B(1)\sim\bin(1,1-a)$ (a Bernoulli
with
$a=\mathbb{P}\left\{B(1)=0\right\}$),
which are also independent from $Y$, namely
\begin{equation*}
    Z_a=B(1)\,Z
\end{equation*}
In summary, given two exponential \rv's $Y\sim\erl_1(\lambda)$
and $Z\sim\erl_1(\lambda)$, and a Bernoulli
$B(1)\sim\bin(1,1-a)$ (all mutually
independent) the \rv
\begin{equation*}
    X= aY+B(1)Z
\end{equation*}
is again an exponential $\erl_1(\lambda)$ defined as the weighed sum
of $Y$ and $Z$: while $a$ is the \emph{deterministic} weight of $Y$,
the weight of $Z$ is \emph{random} and is represented by another
(independent from both $Y$ and $Z$) $0\,$-$1$ Bernoulli \rv\
$B(1)\sim\bin(1,1-a)$. In other
words $X$ is nothing else than the exponential $Y$ \emph{down}
$a$-\emph{rescaled}, plus another independent, but
\emph{intermittent} with frequency $1-a$, exponential $Z$. The
self-decomposability of the exponential laws ensures then that, if
both the parameters of $Y$ and $Z$ are $\lambda$, also $X$
marginally is an $\erl_1(\lambda)$ for every $0<a<1$. It is apparent
on the other hand that, by construction, $X$ and $Y$ are not
independent and it is easy to show
that $a$ represents their correlation coefficient.

Remark that if $Y'\sim\erl_1(\mu)$, then $\alpha
Y'\sim\erl_1\left(\frac{\mu}{\alpha}\right)$ for every $\alpha>0$,
and hence in particular
\begin{equation*}
    Y\eqd\frac{\mu}{\lambda}\,Y'\sim\erl_1(\lambda)
\end{equation*}
As a consequence we could also state the self-decomposability by
means of \emph{exponential} \rv's \emph{with different parameters}
$X\sim\erl_1(\lambda)$ and $Y'\sim\erl_1(\mu)$ because of course we
have
\begin{equation*}
    X= aY+Z_a=\frac{a\mu}{\lambda}\,Y'+Z_a
\end{equation*}
provided that $0<a<1$.

As initially suggested in Iyer et al. \cite{IMM2002}, we take now a sequence of
\iid\ \rv's
\begin{equation*}
    X_k=aY_k+B_k(1)Z_k\qquad\quad k=1,2,\ldots
\end{equation*}
in such a way that for every $k$: $X_k,Y_k,Z_k$ are
$\erl_1(\lambda)$, $B_k(1)$ is
$\bin(1,1-a)$, and $Y_k,Z_k,B_k(1)$
are mutually independent. Add moreover $X_0=Y_0=Z_0=0,\;\PqoP$ to
the list, and then define the point processes
\begin{eqnarray*}
  T_n &=& \sum_{k=0}^nX_k\sim\erl_n(\lambda)\qquad\quad n=0,1,2,\ldots \\
  S_n &=& \frac{\lambda}{\mu}\sum_{k=0}^nY_k\sim\erl_n(\mu)\qquad\quad n=0,1,2,\ldots
\end{eqnarray*}
where $\erl_n(\lambda)$ are Erlang (gamma) laws with \pdf's and
\chf's
\begin{equation*}
    f_n(x)=\lambda\frac{(\lambda x)^{n-1}}{(n-1)!}e^{-\lambda
    x}\mathbbm{1}_{x\ge 0}\qquad\quad\varphi_k(u)=\left(\frac{\lambda}{\lambda-iu}\right)^n\qquad\quad n=0,1,2,\ldots
\end{equation*}
where it is understood that $\erl_0=\bm\delta_0$. We will finally
denote with $N(t)\sim\poiss(\lambda t)$ and $M(t)\sim\poiss(\mu t)$
the \emph{dependent} Poisson processes associated respectively to
$T_n$ and $S_n$, and for our purposes we are interested in finding
an explicit form of
\begin{equation*}
    p_{m,n}(t)=\mathbb{P}\left\{M(t)=m,\,N(t)=n\right\} \qquad\quad
    n,m=0,1,2,\ldots\qquad t\ge0
\end{equation*}
To this end we first introduce the parameter
\begin{equation*}
    \gamma=\frac{a\mu}{\lambda}
\end{equation*}
and the shorthand notations
\begin{eqnarray*}
      \pi_k(\alpha)&=& e^{-\alpha}\frac{\alpha^k}{k!}\qquad\quad k=0,1,\ldots\\
      \beta_\ell(n) &=& \binom{n}{\ell}a^{n-\ell}(1-a)^\ell\qquad\quad \ell\le n=0,1,\ldots
\end{eqnarray*}
respectively for the distributions of a Poisson
$\poiss(\alpha)$, a binomial
$\bin(n,1-a)$ (it is understood that $\beta_0(0)=1$) and a binomial
mixture of shifted Erlang laws $\erl_\ell(\lambda)$, and then we
prove (see Cufaro Petroni and Sabino \cite{cs15}) the following
result
\begin{prop}\label{prop:case1}
If $\gamma \ge 1$, namely $a\mu \ge\lambda $, we have
\begin{eqnarray*}
  p_{m,n}(t)&=&\left\{\begin{array}{ll}
                     0 & \qquad n>m\ge0\\
                     Q_{n,n}(t) & \qquad m=n\ge0 \\
                     Q_{m,n}(t)-Q_{m,n+1}(t) & \qquad m>n\ge0
                   \end{array}
                 \right. \\
  Q_{m,n}(s,t) &=&\sum_{k=n}^m(-1)^k\sum_{j=k}^m\binom{j}{k}\frac{\pi_{m-j}(\mu t)}{(-a)^j}\sum_{\ell=0}^n\beta_\ell(n)\pi_{j+\ell}(\lambda t)\Phi(j+1;j+\ell+1;\lambda
   t)
\end{eqnarray*}
When $\gamma \le 1$, namely $a\mu \le\lambda$, we have
\begin{equation*}
    p_{m,n}(t)=\left\{\begin{array}{ll}
                    \! A_{m,n}(t)-A_{m,n+1}(t)+B_{m,n}(t)-B_{m,n-1}(t) &
                     \quad n>m\ge0 \\
                     \! A_{n,n}(t)-A_{n,n+1}(t)+B_{n,n}(t)+C_{n,n}(t) & \quad m=n\ge0 \\
                     \! A_{m,n}(t)-A_{m,n+1}(t)+C_{m,n}(t)-C_{m,n+1}(t) & \quad m>n\ge0
                   \end{array}
                 \right.
\end{equation*}
where we define for every $n,m\ge0$
\begin{eqnarray*}
      A_{m,n}(t)\!\! &=&\!\pi_m(\mu t)\sum_{k=0}^n\beta_k(n)\left[1+\pi_k(\lambda t-a\mu t)-\sum_{j=0}^k\pi_j(\lambda t-a\mu
      t)\right]
\end{eqnarray*}
while for $n\ge m\ge 0$, and $\lambda t-a\mu t=w$ for short, it is
\begin{eqnarray*}
      B_{m,n}(t) &=&\!\pi_m(\mu
      t)\sum_{k=0}^{n-m}\pi_k\left(\frac{w}{a}\right)\sum_{\ell=0}^{n+1}\beta_\ell(n+1)\frac{w^\ell
      k!}{(k+\ell)!}\,\Phi\left(\ell,k+\ell+1,\frac{1-a}{a}w\right)
\end{eqnarray*}
and for $m\ge n\ge 1$ it is (for $n=0$ we have $C_{m,0}(t)=0$)
\begin{eqnarray*}
  C_{m,n}(t) &=&\!\frac{e^{-(1-a)\mu
  t}}{a^m}\sum_{\ell=1}^n\beta_\ell(n)\sum_{k=n}^m\sum_{j=0}^{\ell-1}\binom{k+\ell-j-1}{k}\\
  &&\qquad (-1)^{\ell-1-j}\pi_j(\lambda t)\pi_{m+\ell-j}(a\mu
  t)\Phi(k+\ell-j,m+\ell-j+1,a\mu t)
\end{eqnarray*}
and
$\Phi(j+1;j+\ell+1;\lambda t)$ for $0\le\ell\le n\le j\le m$ are the confluent hypergeometric functions that are in
fact elementary functions as proved Cufaro Petroni and Sabino \cite{cs15}.
\end{prop}
 \indim
See Cufaro Petroni and Sabino \cite{cs15} including the  more general case for $
    p_{m,n}(s,t)=\mathbb{P}\left\{M(s)=m,\,N(t)=n\right\}
    n,m=0,1,2,\ldots\qquad t\ge0$. Remark that in the
boundary case $\bm{\gamma = 1}$ the previous two expressions
coherently return the same result.
 \findim
\section{The Market Models}\label{sec:The Models}
In this section we adapt the model described in Section \ref{sec:bi_poisson} to the financial context. We consider an usual Black-Scholes (BS) market and a market with geometric Ornstein Uhlenbeck (GOU) processes with jumps similar to the one adopted by Cartea and Figueroa \cite{CarteaFigueroa}. Finally we focus on the Schwartz-Smith model with double jumps that can be seen as a complete cointegrated model with jumps.

Hereafter,  compared to Section \ref{sec:bi_poisson}, $N_1(t)$ and $N_2(t)$  replace  $M(t)$ and $N(t)$ and $\lambda_1$ and $\lambda_2$ replace $\mu$ and $\lambda$, respectively.
\subsection{The GBM plus Jumps Case}\label{subsec:GBMOU}
Consider a BS market with two risky underlying assets whose dynamics are driven by SDEs with the following solution (Merton model):
\begin{equation}\label{eq:gbm_market}
    S_i(T) = \exp\left[
    \log S_i(0) + \left(\mu_i -\frac{1}{2}\sigma_i^2\right)T + \sigma_i W_i(T)+ \sum_{n_i=1}^{N_i (T)}\log J^{n_i}_{i}\right], \quad i=1,2,
    \end{equation}
with $dW_1(t)dW_2(t)=\rho^{(W)} dt$ and log-normal jumps:
\begin{equation}
    J_i = M_i\exp\left(-\frac{\nu_i^2}{2} + \nu_i Z_i\right), \quad i=1,2.
\end{equation}
where $Z_i\sim N(0,1)$ and $Corr(Z_1Z_2)=\rho^{(D)}$. We assume that the compound Poisson processes and BM are independent.

We now concentrate on the logarithm:
\begin{eqnarray}
    \log S_i(T) &\stackrel{d}{=}&  \log S_i(0) + \left(\mu_i -\frac{1}{2}\sigma_i^2\right)T + \sigma_i W_i(T)+ N_i (T) \log M_i \nonumber\\
    &&- \frac{\nu_i^2}{2}N_i (T)+ \nu_i \sum_{n_i=1}^{N_i (T)} Z^{n_i}_{i}, \quad i=1,2.
\end{eqnarray}
The equations above can be rewritten as:
\begin{eqnarray}\label{eq:gbm_jumps}
    \log S_i(T) &\stackrel{d}{=}&  \log S_i(0) + \left(\mu_i -\frac{1}{2}\sigma_i^2\right)T + N_i (T) \log M_i \nonumber\\
    &&- \frac{\nu_i^2}{2}N_i (T) + \sqrt{\sigma_i^2T + N_i (T) \nu_i^2}H_i, \quad i=1,2,
\end{eqnarray}
where given $n$ and $m$, $(H_1,H_2)\sim N\left( \left(
                              \begin{array}{c}
                                0 \\
                                0 \\
                              \end{array}
                            \right),
                            \left(
                              \begin{array}{cc}
                                1, \quad\rho^{(J),n,m} \\
                                \rho^{(J),n,m}, \quad 1 \\
                              \end{array}
                            \right)\right) $
The calculation of $\rho^{(J),n,m}$ can be found in the Appendix \ref{app:rho}.

For simplicity we denote
\begin{equation}\label{eq:var_GBM}
    v_i^{(J,n)}(T) = \left(\sigma_i^{(J,n)}\right)^2=\sigma_i^2T +n\nu_i^2 =v_i^{(C)}(T)+v_i^{(D,n)},
\end{equation}
where $v_i^{(C)}$ and $v_i^{(D)}$ denote the terminal variances of the continuous and discontinuous parts. In case the continuous part of the SDE has a time-dependent volatility function, it is easy to see that the formulas still hold by replacing $v_i^{(C)}(T)$ by $\int_0^T\sigma_i^2(s)ds$.

No-arbitrage conditions imply (see Joshi \cite{Joshi2005} pag 344):
\begin{equation}\label{eq:risk_neutral}
    \mu_i - r = - \lambda_i \mathbb{E}[J_i-1] \quad i=1,2.
\end{equation}

\subsection{The Ornstein-Uhlenbeck  plus Jumps Case}\label{subsec:GOU}
Energy markets often display mean-reversion and jumps. We here consider a one-factor model plus jumps similar to the one introduced in Cartea and Figueroa \cite{CarteaFigueroa}.
Consider a market driven by a stochastic process whose solution is:
\begin{equation}\label{eq:spot_OU}
    S_i(t)=F_i(0,t)\exp\left\{U_i(t) + h(t)\right\}, \quad i=1,2,
\end{equation}
where  $h(t)$ is a pure deterministic function and $U_i(t)$ is
\begin{equation}\label{eq:OU_Jump}
    U_i(t) = U_i(0)e^{-k_it} + \sigma_i\int_0^te^{-k_i\left(t-s\right)}dW_i(s) + e^{-k_it}\sum_{n_i=1}^{N_i(t)}Y^{n_i}_{i}=U_i^{C}(t)+U_i^{D}(t)
\end{equation}
whose SDE is:
\begin{equation}\label{eq:SDA_OU_Jump}
    dU_i(t) = -k_iU_i(t)dt + \sigma_idW_i(t) + e^{-k_it}Y_idN_i(t).
\end{equation}
$Y^{n_i}_{i}$ are copies of $Y_{i}\sim N(M_i,\nu_i^2)$ and $Corr(Y_1,Y_2)=\rho^{(D)}$. Remark that compared to the GBM case the \rv's $Y^{n_i}_{i}$ are not in terms of logarithms and that our model assumes that the size of the jumps decreases in time.
The spot SDE is slightly different from the one adopted in Cartea and Figueroa \cite{CarteaFigueroa}, indeed the exponential term that multiplies the jump component is chosen such that the solution has no random jumps with time-dependent jump size.
Should we have considered as Cartea and Figueroa \cite{CarteaFigueroa},
\begin{equation*}
    dU_i(t) = -k_iU_i(t)dt + \sigma_idW_i(t) + Y_idN_i(t),
\end{equation*}
the solution would have been
\begin{equation*}
    U_i(t) = U_i(0)e^{-k_it} + \sigma_i\int_0^te^{-k_i\left(t-s\right)}dW_i(s) + e^{-k_it}\sum_{n_i=1}^{N_i(t)}Y^{n_i}_{i}e^{k_iT^{n_i}_{i}}
\end{equation*}
where $T^{n_i}_{i}$ are the jump times. This setting leads to less tractable option formulas as it will be shown here below.

In order to get no-arbitrage conditions, we impose $\mathbb{E}\left[S(T)|\mathcal{F}_t\right] = F(t,T)$ and for simplicity we look at  $\mathbb{E}[S(T)] = F(0,T)$ to adjust our parameters and functions.
We then need to compute $\mathbb{E}\left[e^{U_i^{C}(t)+U_i^{D}(t)}\right] = \mathbb{E}\left[e^{U_i^{C}(t)}\right]\mathbb{E}\left[e^{U_i^{D}(t)}\right]$.

It is well known that:
\begin{equation}
    \mathbb{E}\left[e^{U_i^{C}(t)}\right]=\exp{\left(\mathbb{E}\left[U_i^{C}(t)\right]-\frac{1}{2}\mathbb{V}ar\left[U_i^{C}(t)\right]\right)}=e^{a_i(t)}.
\end{equation}
with:
\begin{eqnarray}
    \mathbb{E}\left[U_i^{C}(t)\right]&=&U(0)e^{-k_it},\nonumber \\
    \mathbb{V}ar\left[U_i^{C}(t)\right]&=&\frac{\sigma_i^2}{2k_i}\left(1-e^{-2k_it}\right).
\end{eqnarray}
Hereafter we will assume that $U_i(0)=0, i=1,2$ that does not change the applicability of the model.
Finally we need to calculate:
\begin{equation}
    \mathbb{E}\left[e^{U_i^{D}(t)}\right] = \mathbb{E}\left[\exp\left( e^{-k_it}\sum_{n_i=1}^{N_i(t)}Y^{n_i}_{i}\right)\right]=e^{b_i(t)}.
\end{equation}
Knowing the moment-generating function of the compound Poisson process:
\begin{equation}
    \phi(u) = \mathbb{E}\left[\exp\left\{u\sum_{n=1}^{N_i(t)}Y^{n_i}_{i}\right\}\right] = \exp\left\{\lambda_it\left(\phi_{Y_i}(u)-1\right)\right\}
\end{equation}
where
\begin{equation}
    \phi_{Y_i}(u) = \exp\left\{M_iu+\frac{1}{2}\nu_i^2u^2\right\}
\end{equation}
we easily obtain the required expected value.
\begin{equation}
    \mathbb{E}\left[e^{U_i^{D}(t)}\right] = \phi\left(e^{-k_it}\right).
\end{equation}
\begin{equation}
    b_i(t) = \lambda_it\left(e^{e^{-k_it}\left(M_i+\frac{1}{2}e^{-k_it}\nu_i^2\right)}-1\right)
\end{equation}
Based on the results above, no-arbitrage is given by $h_i(t) = -a_i(t)-b_i(t)$.
The equations for the spot dynamics above can be rewritten as:
\begin{eqnarray}
    \log S_i(t) &\stackrel{d}{=}&  \log F_i(0,t) - b_i(t) + N_i(t)M_ie^{-k_it} + \frac{1}{2}\nu_i^2e^{-2k_it}N_i (t) \nonumber\\ &&-\frac{1}{2}\left(\mathbb{V}ar\left[U_i^{(C)}(t)\right]+\nu_i^2e^{-2k_it}N_i (t)\right)+\nonumber\\
    && \sqrt{\mathbb{V}ar\left[U_i^{(C)}(t)\right] + \nu_i^2e^{-2k_it}N_i(t)}H_i.
\end{eqnarray}
Where $H_i,i=1,2$ have been defined in the previous section.

\subsection{The Schwartz-Smith  plus Jumps Case}\label{subsec:SSM}
Consider the two factor Schwartz-Smith model (see Schwartz Smith \cite{SchwSchm00}):
\begin{eqnarray}\label{eq:SSM1}
    U_1(t) &=& U_1(0)e^{-kt} + \sigma_1\int_0^te^{-k\left(t-s\right)}dW_1(s) + e^{-kt}\sum_{n_1=1}^{N_1(t)}Y^{n_1}_{1}\nonumber\\
    U_2(t) &=& U_2(0) + \mu t + \sigma_2W_2(t)+ \sum_{n_2=1}^{N_2(t)}Y^{n_2}_{2}\nonumber\\
    U(t) &=& U_1(t) + U_2(t).
\end{eqnarray}
where $S(t)=F(0,t)e^{h(t)+U(t)}$ and we assume that the jumps of both process share the same distribution $Y_1,Y_2\sim N(M,\nu)$. Simply taking the differential and some algebra:
\begin{eqnarray}\label{eq:SSM1_SDE}
    dU(t) &=& -k\left(\mu + U_2(t) - U(t)\right)dt + \sigma dW + Y\left(e^{-kt}dN_1 + dN_2\right)
\end{eqnarray}
Should we consider the OU plus compound Poisson as in Cartea and Figueroa \cite{CarteaFigueroa} the process $U(t)$ would be:
\begin{eqnarray}\label{eq:SSM1_CarteaFRig}
    dU(t) &=& -k\left(\mu + U_2(t) - U(t)\right)dt + \sigma_1dW_1 + \sigma_2dW_2 + Y\left(dN_1 + dN_2\right)\nonumber\\
     &=& -k\left(\mu + U_2(t) - U(t)\right)dt + \sigma dW + YdN(t).
\end{eqnarray}
where $\sigma^2 = \sigma_1^2 + \sigma_2^2 + 2\sigma_1\sigma_2\rho^{(W)}$.
With the latter equation above (\ref{eq:SSM1_CarteaFRig}), the log of the spot process can be expressed in terms of one BM and a compound Poisson-like process.

With the same procedure outlined in the previous subsection, no arbitrage conditions can be obtained by taking the (conditional) expectation of the spot process:
\begin{eqnarray}
    \mathbb{E}\left[e^{U^{C}(t)}\right]&=&\mathbb{E}\left[e^{ U_1(0)e^{-kt} + \sigma_i\int_0^te^{-k\left(t-s\right)}dW_1(s) + U_2(0) + \mu t + \sigma_2W_2(t) }\right]\nonumber\\
    &=&\exp{\left(\mathbb{E}\left[U^{C}(t)\right]-\frac{1}{2}\mathbb{V}ar\left[U^{C}(t)\right]\right)}=e^{a(t)}.
\end{eqnarray}
Assuming once more $  U_1(0)=0$ and $U_2(0)=0$ we have:
\begin{eqnarray}
    \mathbb{E}\left[U^{C}(t)\right]&=&\mu t,\nonumber \\
    \mathbb{V}ar\left[U^{C}(t)\right]&=&\frac{\sigma_1^2}{2k}\left(1-e^{-2kt}\right)+\sigma_2^2t +
    \frac{2\rho\sigma_1\sigma_2}{k}\left(1-e^{-kt}\right).
\end{eqnarray}
For the discontinuous component we have:
\begin{eqnarray}
    e^{b(t)} &=& \mathbb{E}\left[e^{U^{D}(t)}\right] = \mathbb{E}\left[\exp\left( e^{-kt}\sum_{n_1=1}^{N_1(t)}Y^{n_1}_{1}+\sum_{n_1=1}^{N_2(t)}Y^{n_2}_{2}\right)\right]= \nonumber\\
     &=& \sum_{m_1,m_2=0}^{+\infty} p_{m_1,m_2}\left[\exp\left( e^{-kt}\sum_{n_1=1}^{m_1}Y^{n_1}_{1}+\sum_{n_1=1}^{m_2}Y^{n_2}_{2}\right)\right]= \nonumber\\
      &=& \sum_{m_1,m_2=0}^{+\infty}p_{m_1,m_2}\phi_Y\left(e^{-kt}\right)^{m_1}\times\phi_Y\left(1\right)^{m_2}
\end{eqnarray}
As done in Section \ref{subsec:GOU}, no-arbitrage is given by $h(t) = -a(t)-b(t)$ where $b(t)$ can be computed numerically.

In contrast, in the case of  Cartea and Figueroa \cite{CarteaFigueroa}, we have:
\begin{eqnarray}
    e^{b(t)} &=& \mathbb{E}\left[e^{U^{D}(t)}\right] = \mathbb{E}\left[\exp\left( e^{-kt}\sum_{n_1=1}^{N_1(t)}Y^{n_1}_{1}e^{kT^{n_1}}+\sum_{n_1=1}^{N_2(t)}Y^{n_2}_{2}\right)\right]
\end{eqnarray}
that is more complex to treat

After some algebra, the log-spot dynamics above can be rewritten as:
\begin{eqnarray}
    \log S_i(t) &\stackrel{d}{=}&  \log F(0,t) - b(t) + \mu t + N_1(t)e^{-kt}\left(M+e^{-kt}\frac{\nu^2}{2}\right) + N_2(t)\left(M+\frac{\nu^2}{2}\right) - \nonumber\\ &-&\frac{1}{2}\left\{\mathbb{V}ar\left[U^{(C)}(t)\right]+\left(e^{-2k_it}N_1 (t)+N_2(t)\right)\right\}\nu^2+ \nonumber\\
    &+&\sqrt{\mathbb{V}ar\left[U^{C}(t)\right] + \left(e^{-2k_it}N_1(t)+N_2(t)\right)\nu^2}\epsilon.
\end{eqnarray}
\section{Risk-neutral Pricing Formulas}\label{sec:rn_pricing}

\subsection{The simple European Plain Vanilla Options Case}\label{subsec:rn_call}

In order to simplify the calculation and the notation, we represent the price of a call option at time zero $c(0)$ in terms of an abstract BS formula
\begin{equation}\label{eq:BS_abstract}
    c(0) = BS\left(P_0, K, r, T, v, q\right).
\end{equation}
Where $P_0$, $K$, $r$, $T$, $v$, $q$ denote the arguments: initial price, strike, risk-free rate, maturity, terminal variance and dividend yield, respectively,  for the Black-Scholes formula. Following the procedure in Joshi \cite{Joshi2005} for the Merton model with jumps, getting a vanilla option pricing formula will then be a matter of plugging the terminal variance, the initial price and the dividend yield into the the abstract formula after applying a conditioning argument on the Poisson probabilities.

\begin{itemize}
    \item \textbf{GBM Case}. We need to rearrange Equation (\ref{eq:gbm_jumps}) given the market of Equation (\ref{eq:gbm_market}) such that we can apply the abstract BS formula for the GBM-plus-jumps case:
    \begin{equation}
        \log S_i(T) = \log S_i(0)+N_i(T)\log M_i + \lambda_i(1-M_i)T -\frac{v_i^{(J,N_i(T))}}{2}+\sqrt{v_i^{(J,N_i(T))}}H_i
    \end{equation}

    The price of a call (put) option on underlying asset $i=1$ given the GBM-plus-jumps market of Equation (\ref{eq:gbm_market}) is then:
    \begin{equation}
        c(0) = \sum_{n=0}^{\infty}\pi_{n_1}(\lambda_1T)BS(S_1^{(n)}(0),K,r,T,v_1^{(J,n)}(T),0).
    \end{equation}
    where
    \begin{equation}\label{eq:bs_spot_call}
        S_1^{(n)}(0)=S_1(0)M_1^n\exp\left[\lambda_1T(1-M_1)\right].
    \end{equation}
    and $v_1^{(J,n)}(T)$ is defined in Equation (\ref{eq:var_GBM}). Please note that the formula here is slightly different from the one in Joshi \cite{Joshi2005} pag. 346, because of our definition of the argument of the abstract BS formula.

    \item \textbf{GOU Case}. In contrast for the OU-plus-jumps market of Equation (\ref{eq:spot_OU}), the starting point argument for the abstract BS formula is:
    \begin{equation}\label{eq:bs_spot_call}
        S_1^{(n)}(0)=F_1(0,T)e^{p_i^n(T)},
    \end{equation}
    where $v_1^{(J,n)}(T)= \mathbb{V}ar\left[U_1^{(C)}(T)\right] + n e^{-2k_1T}\nu_1^2$ and
    \begin{equation}
        p_1^n(t) = -b_1(t) + ne^{-k_1t}\left(\frac{1}{2}e^{-k_1t}\nu_1^2 + M_1\right).
    \end{equation}
    \item \textbf{Schwartz-Smith Case}. Assuming Equation (\ref{eq:SSM1}) a semi-closed form formula can be found following the procedure outlined in the GBM and GOU cases.
        \begin{equation}
            c(0) = \sum_{n_1,n_2=0}^{\infty}\mathbb{P}\left(N_1(T)=n_1, N_2(T)=n_2\right)BS(S^{(n_1,n_2)}(0),K,r,T,v^{(J,n_1,n_2)}(T),0).
        \end{equation}
        where
        \begin{equation}
            S^{(n_1,n_2)}(0)=F(0,t)e^{ -b(t) + \mu t + n_1(t)e^{-kt}\left(M+e^{-kt}\frac{\nu^2}{2}\right) + n_2(t)\left(M+\frac{\nu^2}{2}\right)},
        \end{equation}
        and
        \begin{equation}
            v^{(J,n_1,n_2)}(T) = \mathbb{V}ar\left[U^{C}(T)\right] + \left(e^{-2k_iT}n_1+n_2\right)\nu^2
        \end{equation}
        Moreover, for both settings in Equations (\ref{eq:SSM1}) and (\ref{eq:SSM1_CarteaFRig}) given the no-arbitrage conditions, the simple algorithm to generate the dependent Poisson processes of Section \ref{sec:bi_poisson} offers an easy Monte Carlo implementation to compute the price of vanilla options.
\end{itemize}
Having obtained the risk-neutral conditions on each underlying asset, it is straightforward to obtain formulas for spread options.

\subsection{Spread Options Case}\label{subsec:rn_call}
The application to spread options is the native framework to compare our approach with cointegrated jumps compared to other jump-diffusion cases.
We start considering a spread option with zero-strike, based on the results of Margrabe \cite{Mar78}, with the same conditioning approach applied in Subsection (\ref{subsec:rn_call}) we get that
the price of a spread option with zero strike given the market of Equation (\ref{eq:gbm_market}) or Equation (\ref{eq:spot_OU}) is:
\begin{equation}
    s(0) = \sum_{n_1,n_2=0}^{\infty}\mathbb{P}\left(N_1(T)=n_1;N_2(T)=n_2\right)BS(S_1^{(n_1)}(0),S_2^{(n_2)}(0),0,T,v^{(M,n_1,n_2)}(T),0)
    \footnote{when the Margabe option is written on the spot it can be seen that the price is independent of $r$, in contrast to the situation of a spread option on the forward.},
\end{equation}
where $v^{(M,n_1,n_2)}(T) = v_1^{(J,n_1)}(T) + v_2^{(J,n_2)}(T) - 2\rho^{(J,n_1,n_2)}\sqrt{v_1^{(J,n_1)}v_2^{(J,n_2)}}$ is the spread terminal variance. The definition of $S_2^{m}$ and $v_2^{(J,m)}$ follows from the subsection above.

In the literature different analytical approximations are available  when the strike is not zero (see for instance Deng and Lee \cite{DLZ2006} or Kirk \cite{Kirk95}), the extension to the jump diffusion case is just a matter of adapting the parameters of the approximation. Employing Monte Carlo methods is not complicated because the simulation of the $2$-dimensional path is not a complex task and as well as is the $2$-dimensional Poisson generation.

We will deserve future studies to a market of more than two assets that have a jump component. The current framework cannot cope with multi-asset spread options unless one considers that the third asset has no jump term. Pricing multi-assets spread options then can be tackle via simulation, analytical approximations as done in Deng and Lee \cite{DLZ2010} and Pellegrino and Sabino \cite{PellegrinoSabino_2} or by applying moment matching and using one of the solutions available for two legs as explained in Pellegrino and Sabino \cite{PellegrinoSabino_1}.

In the following, we compare three different Poisson models:
\begin{itemize}
    \item \textbf{Independent Jumps}. $N_1(t)$ and $N_2(t)$ are independent Poisson processes.
    The spread option formula is:
    \begin{equation}\label{eq:spread_ind}
        s(0) = \sum_{n_1,n_2=0}^{\infty}\pi_{n_1}(\lambda_1T)\pi_{n_2}(\lambda_2T)BS(S_1^{(n_1)}(0),S_2^{(n_2)}(0),0,T,v^{(M,n_1,n_2)}(T),0).
    \end{equation}
    \item \textbf{One Common Jump}. $N_i(t)=N(t)+N_i^{X},\,i=1,2$, where $N(t)$ and $ N_i^{X}$ are all mutually independent  Poisson processes.
    The spread option formula is:
    \begin{eqnarray}\label{eq:spread_common}
            s(0)=&&\sum_{n=0,n_1,n_2\ge n}^{\infty}\pi_{n_1-n}(\lambda_{1}^XT)\pi_{n_2-n}(\lambda_{2}^XT)
            \pi_{n}(\lambda T)\times\nonumber\\
            &&BS(S_1^{(n_1-n)}(0),S_2^{(n_2-n)}(0),0,T,v^{(M,n_1-n,n_2-n)},0)
    \end{eqnarray}
    \item \textbf{Cointegrated Jumps}. $N_i(t),\, i=1,2$ described in Subsection \ref{subsec:rn_call}. The spread option formula is:
    \begin{equation}\label{eq:spread_cointegrated}
        s(0) = \sum_{n_1,n_2=0}^{\infty}\mathbb{P}\left(N_1(T)=n_1;N_2(T)=n_2\right)BS(S_1^{(n_1)}(0),S_2^{(n_2)}(0),0,T,v^{(M,n_1,n_2)}(T),0)
    \end{equation}
    where $p_{n_1,n_2}=\mathbb{P}\left(N_1(T)=n_1;N_2(T)=n_2\right)$ are defined in Proposition \ref{prop:case1}.
\end{itemize}
The payoff of the spread options above considers the values of the two underlying at the same time $T$. Other types of spread options instead look at the two underlying at different times, e.g. the payoff may be $\left(S_1(T_1)-S_2(T_2)\right)^+$, $T_2<T_1$. In this case one needs to readapt the formulas and consider the probabilities $p_{n_1n_2}=\mathbb{P}\left(N_1(T_1)=n_1;N_2(T_2)=n_2\right)$ and they can be found in Cufaro Petroni and Sabino \cite{cs15}.

\section{Numerical Experiments}\label{sec:numerical_experiments}
In this section we presents the numerical experiments assuming the GBM and GOU dynamics plus jumps explained in the previous sections.
The case with GBM considers realistic parameters and is meant to study the spread option values with different types of bivariate Poisson processes; in contrast the GOU case is based on real data of EEX and Powernext day-ahead prices.

\subsection{GBM. Application to Spread Options}\label{subsec:GBM}
We compare the spread option value obtained using Equations (\ref{eq:spread_ind})-(\ref{eq:spread_cointegrated}) changing the correlation between the two Poisson processes. In particular, assuming $N_i=N(t)+N_i^{X}(t)$ we have $\mathbb{C}ov[N_1(t),N_2(t)]=\mathbb{V}ar[N(t)]=\lambda t$, then the instantaneous correlation is $\rho_{N_1N_2}=\frac{\lambda}{\sqrt{\lambda_1\lambda_2}}$ and is time-independent; in case of cointegrated jumps this can be obtained numerically. Naturally, in case $\lambda_1\neq\lambda_2$ the perfect correlation cannot be obtained.

We consider two cases:
\begin{enumerate}
    \item[Case A]. $\lambda_1 = \lambda_2 = 20$, where for cointegrated jumps $\gamma=a<1$.
    \item[Case B]. $\lambda_1 = 40$,  $\lambda_2 = 20$ where for cointegrated jumps $\gamma<1$ can assume values lower or higher than $1$, when $a<0.5$ and $a>0.5$, respectively.
\end{enumerate}
The parameters  are shown in Table \ref{tab:Parameters:GBM}. In both cases we consider an at-the-money spread option with zero strike, $K=0$ and maturity $T=1$ such that we can use the exact Margrabe formula. One can use some approximations techniques for the spread option value for non-zero strikes without changing the validity of our tests.

We also compute, see Table \ref{tab:Spread_No_Ind_Jumps}, the spread value with a pure GBM with no jumps with  the parameters of the first column of the first table in Table \ref{tab:Parameters:GBM} that are chosen such that they match the average spread terminal variance $v^{(M, \lfloor\lambda_1 T\rfloor, \lfloor\lambda_2 T\rfloor)}$, where $\lfloor\cdot\rfloor$ denotes the integer part.
\begin{table}
    \centering
    \small
    \caption{Parameters of the GBM and Compound Poisson processes}\label{tab:Parameters:GBM}
    \subtable[Continuous Part.]{
        \begin{tabular}{|c|c|c|c|}
            \hline
            & \multicolumn{2}{c|}{No Jump} & With Jumps\\
            \hline
            & Case A& Case B & \\
            \hline
            $S_1(0)$  &  $100$  &  $100$ &  $100$ \\
            \hline
            $S_2(0)$ & $100$  &  $100$ & $100$\\
            \hline
            $\sigma_1$ & $0.49$ & $0.37$ & $0.2$\\
            \hline
            $\sigma_2$ & $0.35$ & $0.23$ & $0.15$\\
            \hline
            $\rho^{(W)}(\%)$ & $96$ & $60$ & $80$\\
            \hline
        \end{tabular}
    }\quad
    \subtable[Discontinuous Part]{
        \small
        \begin{tabular}{|c|c|c|}
            \hline
            & Case A& Case B\\
            \hline
            $\rho^{(D)}$(\%) & $99$ & $50$\\
            \hline
            $\lambda_1$ & $20$ & $40$\\
            \hline
            $\lambda_2$ & $20$ & $20$\\
            \hline
            $\nu_1$ & $0.10$ & $0.05$\\
            \hline
            $\nu_2$ & $0.07$ & $0.04$\\
            \hline
            $M_1$ & $1.1$ & $1.05$\\
            \hline
            $M_2$ & $1.1$ & $1.05$\\
            \hline
        \end{tabular}
    }
\end{table}

\begin{figure}
    \centering
    \subfigure[Case A]{
        \includegraphics[width=70mm]{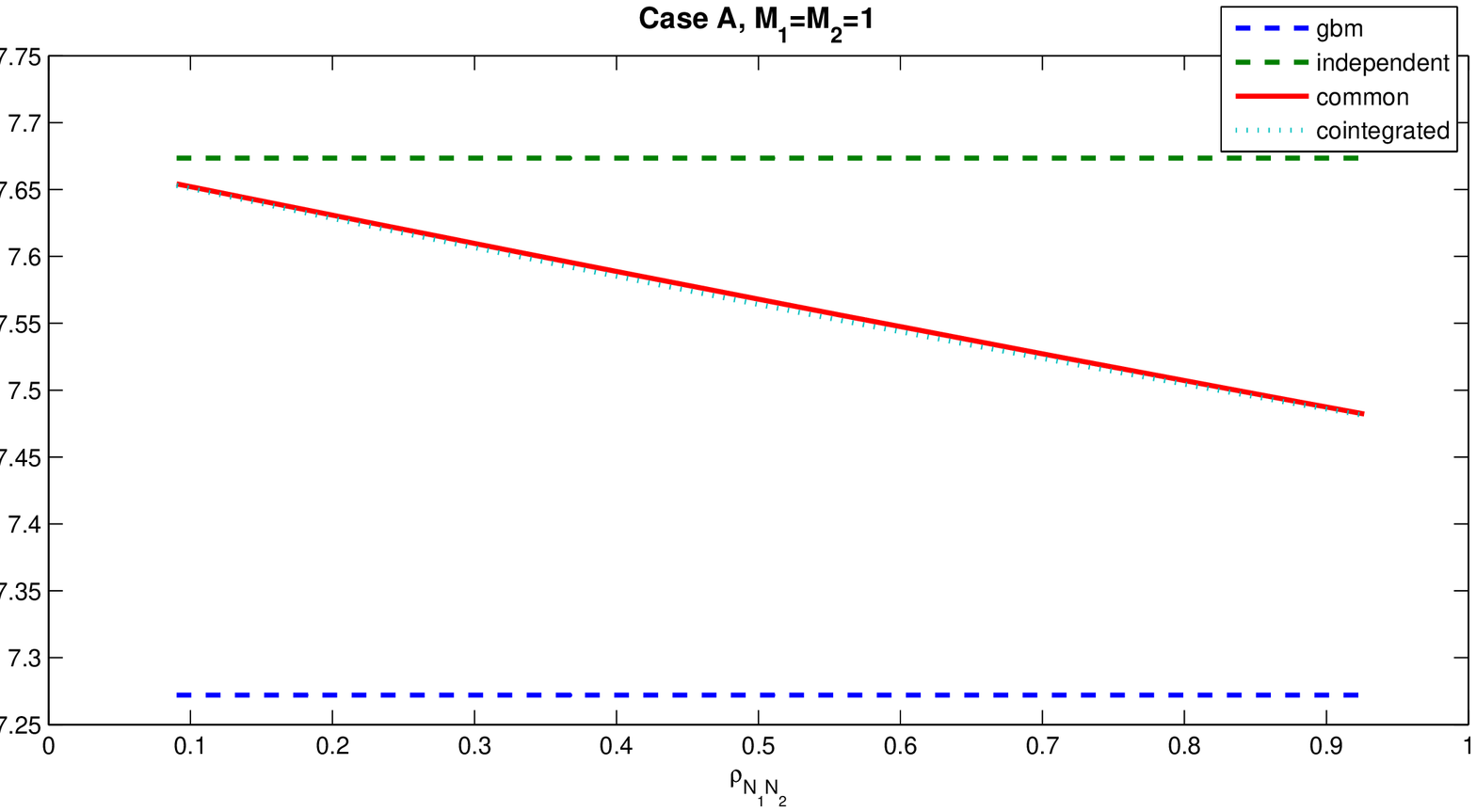}
        }\,
    \subfigure[Case B]{
        \includegraphics[width=70mm]{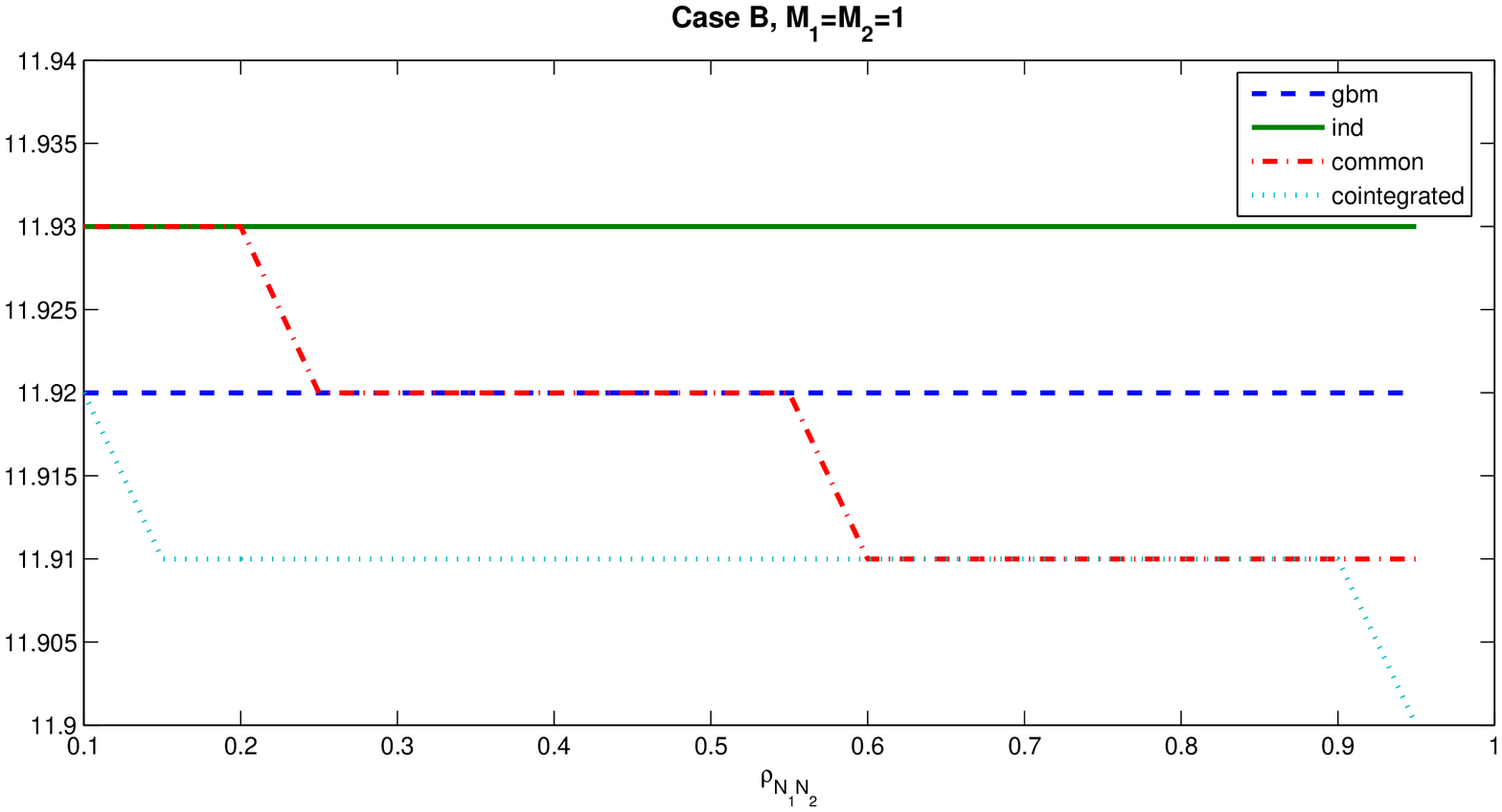}
    }
    \caption{Spread Option Values in the Cases A and B when $M_1=M_2=1$}\label{fig:no_mean_jump}
\end{figure}%

We first observe that the behavior of Equations (\ref{eq:spread_ind})-(\ref{eq:spread_cointegrated}) depends on the values of the probabilities $p_{n_1,n_2}$ and the values of the BS formulas separately. The former quantities do not depend on the distribution of the jumps while the latter ones are independent on the structure of dependence between the Poisson processes.

Figure \ref{fig:no_mean_jump} clearly shows that the expected jump size has a relevant impact in the option value because under the assumption $M_1=M_2=1$ the price in almost independent on the choice of the Poisson model.

\begin{figure}
    \center
    \subfigure[$\lambda_1=\lambda_2=20$]{
    \includegraphics[width=180mm, height=110mm]{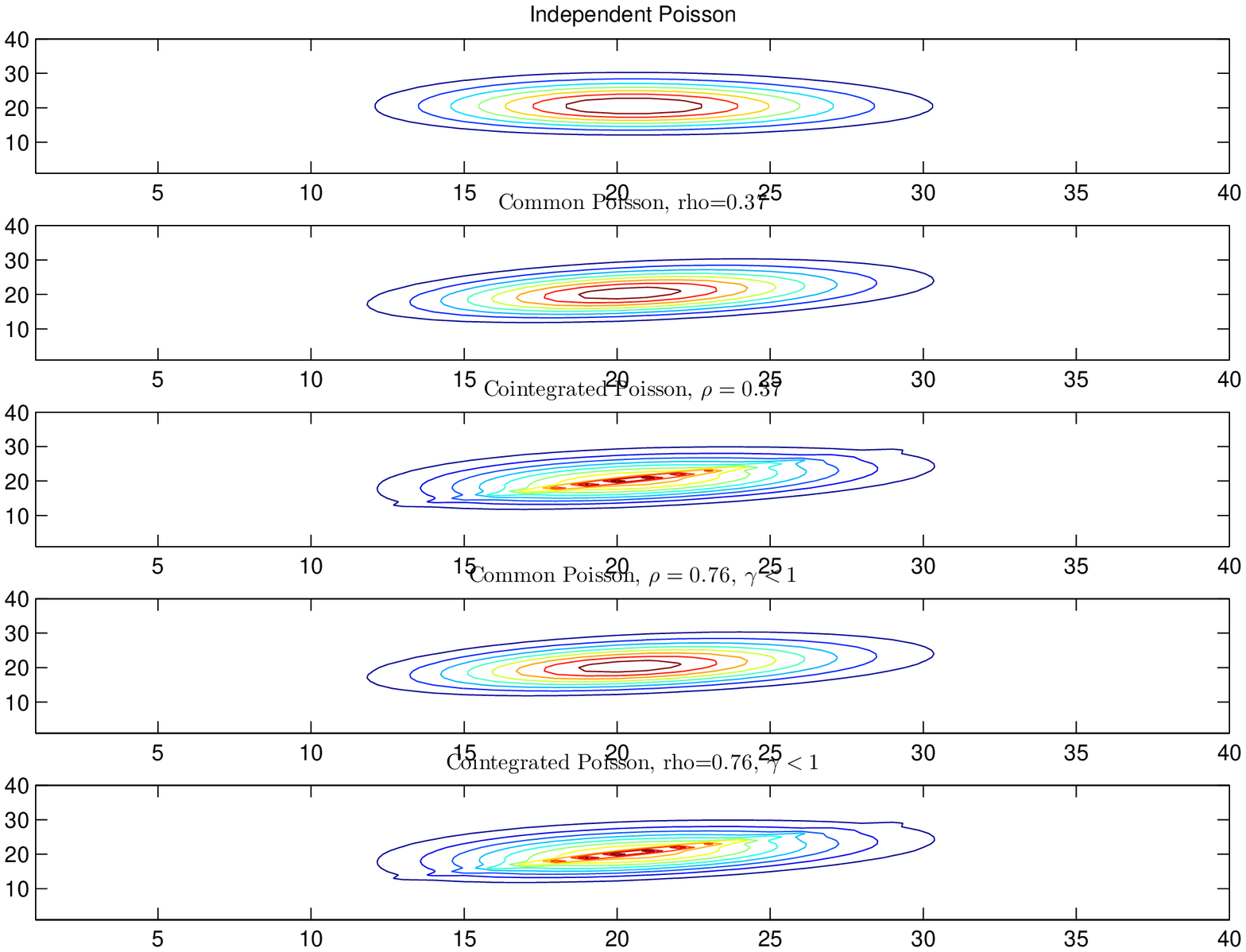}
    }
    \subfigure[$\lambda_1=40$, $\lambda_2=20$]{
    \includegraphics[width=180mm, height=110mm]{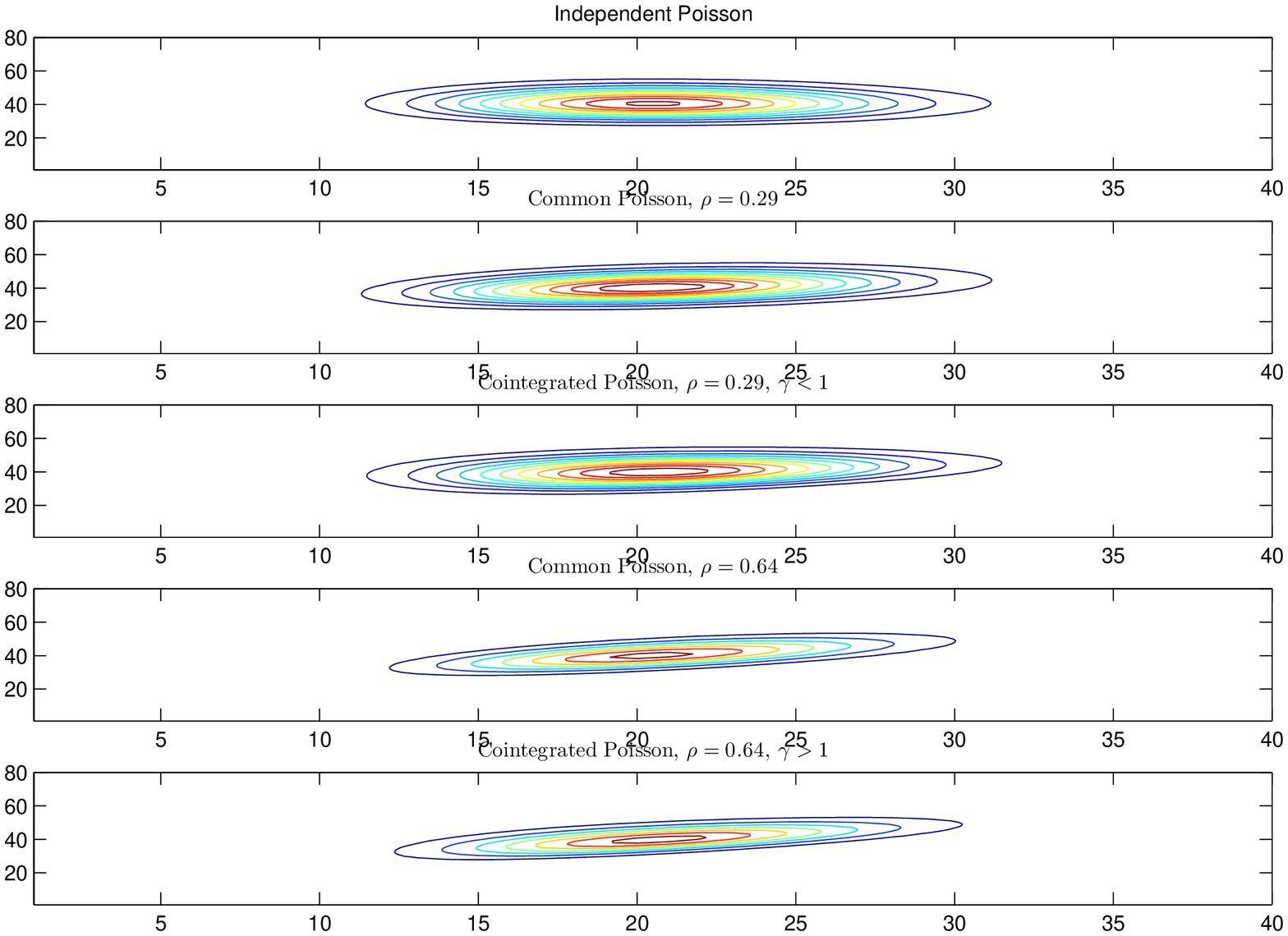}
    }
  \caption{Contour plot of the joint probabilities of the Poisson processes.}\label{fig:Contour}
\end{figure}

Figures \ref{fig:Contour} show the difference among the joint probabilities of the Poisson processes. The isolines of the contour plot of $p_{n_1n_2}$ resemble to a sort of ellipse whose axis are parallel to the X-Y axis that is expected for independent
Poisson both in case A and B. The positive correlation of the Poisson processes in the other configuration is reflected by the fact that the axis are now rotated counterclockwise. In addition for cointegrated Poisson the higher value of the probabilities is more concentrated around the expected value.

\begin{table}
    \centering
    \caption{Spread Option Values without jumps and independent Compound Poisson processes.}\label{tab:Spread_No_Ind_Jumps}
    \small
    \begin{tabular}{|c|c|c|c|c|}
        \hline
    	&\multicolumn{2}{c|}{No Jump} &  \multicolumn{2}{c|}{Independent Jump}\\
        \hline
        & Case A & Case B & Case A & Case B\\
        \hline
        Option Value & $7.27$ & $11.92$ & $25.23$ & $19.27$\\
        \hline
    \end{tabular}
\end{table} 
This is more evident in the figure with $\lambda_1=\lambda_2$ and it is worth noticing once more that for $\gamma>1$ the matrix $p_{n_1n_2}$ is lower triangular for the cointegrated Poisson while it is full for the structure with one common Poisson process.
\begin{table}
    \centering
    \caption{Spread Option Values with Common and Cointegrated Compound Poisson. }\label{tab:Spread_Common_Cointegrated}
    \small
    \begin{tabular}{|c||c|c|c|c||c|c|c|c|}
        \hline
        &\multicolumn{2}{c}{Case A} & \multicolumn{2}{|c||}{Option Value Case A}&\multicolumn{2}{c}{Case B} & \multicolumn{2}{|c|}{Option Value Case B}\\
        \hline
        $a$ & $\rho_{N_1N_2}$(\%) & $\lambda$ & Common & Cointegrated & $\rho_{N_1N_2}$(\%) & $\lambda$ & Common & Cointegrated\\
        \hline
        $0.1$ & $9$ & $1.80$ & $24.30$ & $ 24.22 $ & $7$ & $2.09$ & $18.87$ & $18.87$\\
        \hline
        $0.15$ & $14$ & $2.71$ & $23.76$ & $ 23.64 $ & $11$ & $3.13$ & $18.66$ & $18.67$\\
        \hline
        $0.2$ & $18$ & $3.63$ & $23.20$ & $ 23.05 $ & $15$ & $4.16$ & $18.45$ & $18.46$\\
        \hline
        $0.25$ & $23$ & $4.55$ & $22.63$ & $ 22.44 $ & $18$ & $5.19$ & $18.25$ & $18.26$\\
        \hline
        $0.3$ & $27$ & $5.47$ & $22.04$ & $ 21.81 $ & $22$ & $6.21$ & $18.04$ & $18.05$\\
        \hline
        $0.35$ & $32$ & $6.40$ & $21.42$ & $ 21.16 $ & $26$ & $7.23$ & $17.83$ & $17.83$\\
        \hline
        $0.4$ & $37$ & $7.34$ & $20.78$ & $ 20.48 $ & $29$ & $8.24$ & $17.61$ & $17.62$\\
        \hline
        $0.45$ & $41$ & $8.29$ & $20.11$ & $ 19.78 $ & $33$ & $9.25$ & $17.40$ & $17.40$\\
        \hline
        $0.5$ & $46$ & $9.24$ & $19.41$ & $ 19.05 $ & $36$ & $10.25$ & $17.18$ & $17.18$\\
        \hline
        $0.55$ & $51$ & $10.20$ & $18.68$ & $ 18.29 $ & $40$ & $11.25$ & $16.97$ & $16.96$\\
        \hline
        $0.6$ & $56$ & $11.17$ & $17.90$ & $ 17.49 $ & $43$ & $12.24$ & $16.75$ & $16.74$\\
        \hline
        $0.65$ & $61$ & $12.16$ & $17.08$ & $ 16.64 $ & $47$ & $13.23$ & $16.53$ & $16.51$\\
        \hline
        $0.7$ & $66$ & $13.15$ & $16.20$ & $ 15.74 $ & $50$ & $14.21$ & $16.30$ & $16.29$\\
        \hline
        $0.75$ & $71$ & $14.17$ & $15.25$ & $ 14.78 $ & $54$ & $15.19$ & $16.08$ & $16.06$\\
        \hline
        $0.8$ & $76$ & $15.20$ & $14.21$ & $ 13.75 $ & $57$ & $16.16$ & $15.85$ & $15.83$\\
        \hline
        $0.85$ & $81$ & $16.26$ & $13.06$ & $ 12.61 $ & $61$ & $17.13$ & $15.62$ & $15.60$\\
        \hline
        $0.9$ & $87$ & $17.36$ & $11.74$ & $ 11.33 $ & $64$ & $18.09$ & $15.38$ & $15.37$\\
        \hline
        $0.95$ & $93$ & $18.53$ & $10.14$ & $ 9.82 $ & $67$ & $19.05$ & $15.15$ & $15.14$\\
        \hline
    \end{tabular}
\end{table}

Finally, Table \ref{tab:Spread_Common_Cointegrated} displays the results with expected jump size different from zero. $\lambda$ where chosen such that the correlations of the `cointegrated' and `common' Poisson processes coincides.

The effect of the correlation between the Poisson processes is noticeable: the value of the spread option is decreasing when $\rho_{N_1N_2}$ is increasing that is in line with the intuition because the spread terminal variance decreases.

In the case A, the jump sizes are perfectly correlated and the spread option values using the common Poisson setting is always higher than the values obtained with our methodology. This is somehow reflected by the concentration of the isoline of the probabilities. Using a common Poisson reduces the spectrum of jump events, for instance in an extreme setting where $\lambda=\lambda_1=\lambda_2$, $N_1(t)$ cannot jump more that $N_2(t)$, while this is not the case for the cointegrated Poisson process. In contrast, the choice of the Poisson model has no remarkable effect on the price of the spread option in the configuration B. In the parameters configuration of Table \ref{tab:Parameters:GBM}, for values obtained with the same $\rho_{N_1N_2}$, the price of the spread option seems to highly depend on the number of the jumps of both processes rather than when they occurred and that explains the small differences. Furthermore, assuming for instance $\lambda_2^X=0$ implies $\lambda_2=\lambda$ and $N_2(t)$ cannot have more jumps than $N_2(t)$ that coincides with the properties of our model only if $\gamma>1$, that means that our model gives a richer set of combinations.

\subsection{GOU. Application to Power Interconnectors}\label{subsec:GOU}
In this section we apply our methodology in order to price a power interconnectors between EEX and Powernext  modeled as a spread option (Powernext minus EEX). We assume that each dynamics behaves as a GOU plus a compound Poisson and calibrate the parameters. As done for the GBM example we do not consider transport costs (no strike) and adopt the Margrabe formula in Equations (\ref{eq:spread_ind})-(\ref{eq:spread_cointegrated}).

In particular, the calculation date is end of December 2015 with a historical time window of $2$ years for the estimation period. We concentrate then on the power spread value for the first and second quarters, Q1, Q2 in 2016\footnote{The technique here discussed does not reflect UGC view.}. The choice of the maturities above is justified by the fact that in practice pure spot models are used for short maturities, while it is a common practice to use forward models for longer maturities and add a volatility premium to the forward volatility.

Because the expectations in the pricing formulas are in the risk neutral measure, some observations are necessary for the estimation of the parameters of the spot model. Some approaches propose to assume to live already in the risk-neutral world or to incorporate
a market price of risk in the drift (see Lucia and Schwartz \cite{LS02}). The main goal of our work is not the parameters estimation  and for sake of simplicity we do not consider the market price of risk.

The estimation procedure can be split into two steps. As a first step, after filtering out the time-dependent components of each process, one can estimate the parameters of the one-dimensional processes, $\theta_i=\left(k_i,\sigma_i,\lambda_i,M_i,\nu_i\right)$. As a second step then, one can estimate the remaining joint parameters defined by the two-dimensional model\footnote{An example of how to derive the parameters of the GOU process in Cartea and Figueroa \cite{CarteaFigueroa} can be found at http://de.mathworks.com/help/fininst/simulating-electricity-prices-with-mean-reversion-and-jump-diffusion.html. Although different, the procedure can be adapted to our case.}

Consider an equally spaced time grid $t_0,t_1,\dots,t_T$ with $t_{i+1}-t_i=\Delta t$ and the Euler scheme of each SDE in Equation (\ref{eq:SDA_OU_Jump})
\begin{equation}\label{eq:SDA_OU_Jump}
    U_i(t+1) = \left(1-k_i\Delta t\right)U_i(t) + \sigma_i\sqrt{\Delta t}\epsilon_{i,t+1} +e^{-k_it} \mathbbm{1}_{i}(t+1)Y_i.
\end{equation}
where
\begin{equation}
     \mathbbm{1}_i(t+1) =
        \begin{cases}
            1, & \mbox{with probability } \lambda_i\Delta t\\
            0, & \mbox{with probability } 1-\lambda_i\Delta t.
        \end{cases}
\end{equation}
Hence the transition density is a combination of Gaussian densities:
\begin{equation}
    p_i\left(U_i(t+1),t+1|U_i(t),t\right) = \left(1-\lambda_i\Delta t\right)\mathcal{N}\left(\mu_i^{C}(t),\sigma^{C}(t)\right) + \lambda_i\Delta t\mathcal{N}\left(\mu_i^{J}(t),\sigma^{J}(t)\right).
\end{equation}
$\mathcal{N}(x)$ denotes the density function of a Gaussian random variable and $\mu_i^{C}(t) = \left(1-k_i\Delta t\right)U_i(t)$,  $\mu_i^{J}(t) = \left(1-k_i\Delta t\right)U_i(t) + M_ie^{-k_it}$
and $\sigma_i^{C}(t)= \sigma_i\sqrt{\Delta t} $, $(\sigma_i^{J}(t))^2= \sigma_i^2\Delta t +e^{-2k_it}\nu_i^2$.
The parameters $\theta_i=\left(k_i,\sigma_i,\lambda_i,M_i,\nu_i\right)$ can be calibrated by minimizing the log-likelihood function with the usual constrains on the parameters:
\begin{equation}
    \theta_i=argmin\sum_{t=0}^{T-1}\log\left(p_i\left(U_i(t+1),t+1|U_i(t),t\right)\right).
\end{equation}
The calibration of the parameters for the two-dimensional process depends on the model specification written in Section \ref{sec:bi_poisson}.
\begin{itemize}
    \item \textbf{Independent Jumps}. In case of independent Poisson processes the joint probability are simply:
        \begin{eqnarray}\label{eq:ind_poisson}
            p_{0,0} = \left(1-\lambda_1 \Delta t\right)\left(1- \lambda_2 \Delta t\right) , &\quad& p_{1,0} = \lambda_1\Delta t\left(1- \lambda_2 \Delta t\right)  \nonumber\\
            p_{0,1} = \left(1-\lambda_1 \Delta t\right)\lambda_2\Delta t, &\quad& p_{1,1} = 1 -  p_{0,1} -  p_{1,0} - p_{0,0}.
        \end{eqnarray}
        The only two remaining parameters to estimate are $\rho^{(W)}$ and $\rho^{(J)}$ and can be obtained by minimizing the log-likelihood of the two dimensional process. The transition density is
        \begin{eqnarray}\label{eq:transition_2dim}
            &&p\left(U(t+1),t+1|U(t),t\right) = \nonumber\\
            &&\mathcal{N}\left(\mu^{CC}(t),\Sigma^{CC}(t)\right)p_{0,0} +
            \mathcal{N}\left(\mu^{CJ}(t),\Sigma^{CJ}(t)\right)p_{0,1} +
            \nonumber\\
            &&\mathcal{N}\left(\mu^{JC}(t),\Sigma^{JC}(t)\right)p_{1,0}+  \mathcal{N}\left(\mu^{JJ}(t),\Sigma^{JJ}(t)\right)p_{1,1}.
        \end{eqnarray}
        where
        \begin{eqnarray*}
            \mu^{CC}(t)=\left(\mu_1^{C}(t),\mu_2^{C}(t)\right)&,&
                \Sigma^{CC}(t)=\left(
                \begin{array}{cc}
                (\sigma_1^{C})^2& \rho^{(W)}\sigma_1^{C}\sigma_2^{C} \\
                    \rho^{(W)}\sigma_1^{C}\sigma_2^{C} & (\sigma_2^{C})^2
            \end{array}
            \right),
        \end{eqnarray*}
        \begin{eqnarray*}
            \mu^{CJ}(t)=\left(\mu_1^{C}(t),\mu_2^{J}(t)\right)&,&
            \Sigma^{CJ}(t)=\left(
            \begin{array}{cc}
            (\sigma_1^{C})^2& \rho^{(W)}\sigma_1^{C}\sigma_2^{C}\\
                   \rho^{(W)} \sigma_1^{C}\sigma_2^{C} & (\sigma_2^{J})^2
                \end{array}
            \right),
        \end{eqnarray*}
        \begin{eqnarray*}
            \mu^{JC}(t)=\left(\mu_1^{J}(t),\mu_2^{C}(t)\right)&,&
            \Sigma^{JC}(t)=\left(
            \begin{array}{cc}
                (\sigma_1^{J})^2& \rho^{(W)}\sigma_1^{C}\sigma_2^{C} \\
                \rho^{(W)}\sigma_1^{C}\sigma_2^{C} & (\sigma_2^{C})^2
        \end{array}
        \right),
        \end{eqnarray*}
        \begin{eqnarray*}
            \mu^{JJ}(t)\left(\mu_1^{J}(t),\mu_2^{J}(t)\right)&,&
            \Sigma^{JJ}(t)=\left(
            \begin{array}{cc}
                (\sigma_1^{J})^2& \rho^{(J)}\sigma_1^{J}\sigma_1^{J} \\
                    \rho^{(J)}\sigma_1^{J}\sigma_2^{J} & (\sigma_2^{J})^2
            \end{array}
            \right).
        \end{eqnarray*}
        We do not neglect the $o\left(\Delta t^2\right)$ terms that are necessary to estimate $\rho^{(D)}$.
        \item \textbf{Common Jumps}. One cannot detect the presence of the common Poisson process only looking at each log process independently.

            After some algebra, the pair $\left(\mathbbm{1}_i(t+1)\right)$ is bi-dimensional Bernoulli \rv\ with:
            \begin{eqnarray}\label{eq:com_poisson}
                p_{0,0} = 1-\left(\lambda_1^X+\lambda_2^X+\lambda\right) \Delta t , &\quad& p_{1,0} = \lambda_1^X\Delta t  \nonumber\\
                p_{0,1} = \lambda_2^X\Delta t, &\quad& p_{1,1} = \lambda\Delta t = 1 -  p_{0,1} -  p_{1,0} - p_{0,0}.
            \end{eqnarray}
             In contrast to the case above, we neglect $o\left(\Delta t^2\right)$ terms that implies that we are neglecting the possibility that the Poisson processes $N_1^{X}(t)$ and $N_2^{X}(t)$ jump simultaneously in the unit of time $\Delta t$. The functional form of the transition density is the one of Equation (\ref{eq:transition_2dim}), with different probability weights.
        \item \textbf{Cointegrated Jumps}
        \begin{itemize}
        \item Case 1. $\gamma>1$. Based on the results of Proposition (\ref{prop:case1}) up to $O(\Delta ^2)$ terms we have:
            \begin{eqnarray}\label{eq:a_ge_1}
                p_{0,0} =  1- \lambda_1\Delta t, &\quad& p_{1,0} = (\lambda_1-\lambda_2)\Delta t - \lambda_1\left(\frac{\lambda_1}{\gamma}-\lambda_2\right)\Delta t^2\nonumber\\
                p_{0,1} = 0, &\quad& p_{1,1} = \lambda_2\Delta t.
            \end{eqnarray}
        \item Case 2. $0<\gamma \le 1$.
            \begin{eqnarray}\label{eq:a_ge_2}
                p_{0,0} = 1 - \left(\lambda_1 + \lambda_2(1-\gamma)\right)\Delta t , &\quad& p_{0,1} = \lambda_2(1-\gamma)\Delta t\nonumber\\
                p_{1,0} = (\lambda_1-\gamma\lambda_2)\Delta t, &\quad& p_{1,1} = \gamma\lambda_2\Delta t.
            \end{eqnarray}
            Here above we neglect $o(\Delta t^2)$ terms (see Appendix \ref{app:integrals} for the proof of this last case)
        \end{itemize}
\end{itemize}
The results of the calibration are shown in Table \ref{tab:Parameters:GOU}. The expected jump sizes are both negative and their correlation ($\rho^{(D)}\approx 0 $) is very small and we will neglect it hereafter. Comparing the values of $\lambda$ and $a$ or $\gamma$, we can conclude that the correlation between the two Poisson processes is of the the order of $45\%$. In the case of Poisson processes with common jumps, the correlation is not time dependent, while in the cointegration case the correlation depends on time and it is not $a$. However, $a$ can give a reasonable order of magnitude for the correlation.
\begin{table}
    \centering
    \small
    \caption{Market parameters for EEX and Powernext}\label{tab:Parameters:GOU}
    \subtable[Parameters of the Single Underlyings.]{
        \begin{tabular}{|c|c|c|c|c|c|}
            \hline
            Market & $k$ & $\sigma_i$ & $\mu_i$ & $\nu_i$ & $\lambda_i$\\
            \hline
            EEX & $42.50$ & $1.66$ & $-0.10$ & $0.16$ & $95.32$\\
            \hline
            Powernext	& $41.64$	& $1.52$ & $-0.06$ & $0.38$ & $56.74$\\
            \hline
        \end{tabular}
    }\quad
    \subtable[Common Parameters]{
        \small
        \begin{tabular}{|c|c|c|c|c|}
            \hline
            Method & $\rho^{(W)}(\%)$ &  $\lambda$ & $a$\\
            \hline
            Independent &	$43$  & NA & NA\\
            \hline
            Common & $43$ &	$-1.0$ & NA\\
            \hline
            Cointegrated &	$43$ & NA & $0.44$\\
            \hline
        \end{tabular}
    }
\end{table}

\begin{table}
    \centering
    \small
    \caption{Powernext-EEX Interconnector Prices}\label{tab:Parameters:Tranport_Prices}
    \begin{tabular}{|c|c|c|c|}
        \hline
    	& \multicolumn{3}{c|}{Interconnector Value (EUR)} \\
        \hline		
    	& Independent &	Common & Cointegrated \\
        \hline
        Jan & $289.51$ & $289.64$ & $310.08$\\
        \hline
        Feb & $287.40$ & $287.40$ & $290.81$\\
        \hline        
        Mar & $263.17$ & $263.17$ & $263.37$\\
        \hline
        Q2 & $405.05$ & $405.05$ & $405.06$\\
        \hline
    \end{tabular}
\end{table}


Table \ref{tab:Parameters:Tranport_Prices} shows the values of the power-spread in the quarters Q1 and Q2 and the detail of the first three months with the different dynamics. As shown in Section \ref{subsec:rn_call}, the pricing formulas are double sums over probabilities and BS formulas. The three models differ in the Poisson probability terms, while the values of the BS terms are the same, hence the price difference is only attributable to the different bi-dimensional Poisson law.

The results in Table \ref{tab:Parameters:Tranport_Prices} show that the prices of the interconnector are similar for the cases with independent and common jumps for all the maturities. However, the model with cointegrated jumps returns higher values for all the maturities with a larger difference in January and February. The fact that the prices obtained with our methodology are larger than the ones with independent jumps may seem counterintuitive and wrong if one relies on linear correlation only, indeed one would expect the opposite. The cases of common and independent jumps have similar results because the correlation between the Poisson processes of the former Poisson choice is not high and apparently the different probability weights do not have a remarkable effect on the prices. In contrast, compared to the other approaches, our model gives more emphasis to when the jumps occur and not only to how many jumps have occurred. 

A certain news to one of the two markets may cause an instantaneous propagation of the information and simultaneous jump in both spot prices (if the jumps belong to the common source) while our solution has not this restriction. A same shock to one of the two markets may have caused both prices to jumps but at different times and that explains why the values are higher. Finally, we observe that at longer maturities the difference reduces because of the effect of the mean reversion rate.

Once more we remark how our methodology, that is parameterized by $\gamma$ and $a$, the correlation between the exponential \rv's that construct the Poisson process, implies a structure that goes beyond the linear correlation and its effects are remarkable in the examples illustrated above.

\section{Conclusion and Future Studies}\label{sec:Conclusions}
Based on the concept of self-decomposability we have studied the use of the $2$-dimensional co-dependent Poisson processes proposed in Cufaro Petroni and Sabino \cite{cs15} to model energy derivatives and in general to price spread options. Due to the particular relationships among inter arrival times, we can see this dependence as a form of coitegration among jumps.

To put into the context of modeling energy market and facilities, we have shown how to combine $2$-dimensional compound Poisson processes with Geometric Browian Motions and Geometric Ornstein and Uhlenbeck dynamics. In the latter case, we have adopted a dynamics for day-ahead prices that allows (semi-)closed formulas for plain vanilla options with an easy derivation of risk-neutral conditions.

Focusing on the pricing of spread options, we have compared the option prices using our methodology and different types of Poisson processes. We have shown that our methodology can cope with a wide range of possibilities that go beyond the pure correlation between marginal Poisson and can answer several questions that arise in the financial context.

In our study we have considered power interconnectors but the applicability can be extended to other financial situations. Straightforward applications are in credit and insurance risk where our approach can answer questions regarding the time of contagion or time of propagation of certain information.

In addition, the self-decomposability and subordination technique can be promising tools to study dependency beyond the Gaussian-{\ito} world. For instance, in Cufaro Petroni and Sabino \cite{cs15} we have detailed how to obtain dependent Erlang (Gamma) \rv's that can be used to create and simulate dependent variance gamma processes. Furthermore in recent papers, Sexton and Hanzon \cite{SextonHanzon12} and Hanzon et al. \cite{HanzonHollandSexton12} have studied the use of two sided Exponential-Polynomial-Trigonometric (ETP) density functions to option pricing where EPT are distributions with a strictly proper rational characteristic function. Due to the fact that the Erlang and exponential distributions belong to this class, it will be worthwhile to investigate the use of self-decomposability to create dependence for this larger class of distributions.

\newpage
\appendix

\section{Calculation of $\rho^{(J)}$}\label{app:rho}
\begin{itemize}
    \item The GBM plus Jumps Case.
    \begin{eqnarray}
        \rho^{(J),n,m} &=& Corr\left[\sigma_1W_1(T)+\sqrt{n}\nu_1Z_1,\sigma_2W_2(T)+\sqrt{m}\nu_2Z_2\right]=\nonumber \\
         &=& =\frac{\rho^{(W)} \sigma_1\sigma_2T + \rho^{(D)}\sqrt{nm}\nu_1\nu_2}
        {\sqrt{v_1^{(J,n)}(T)v_2^{(J,m)}(T)}}.
    \end{eqnarray}
    \item The Ornstein-Uhlenbeck  plus Jumps Case.
    \begin{eqnarray}
        \rho^{(J),n,m} &=& Corr\left(L_1,L_2\right)=
        \nonumber \\
         &=&\frac{\frac{\rho^{(W)} \sigma_1\sigma_2}{2\sqrt{k_1k_2}}\sqrt{1-e^{-2k_1t}}\sqrt{1-e^{-2k_2t}} + \rho^{(D)}\sqrt{nm}\nu_1\nu_2e^{-(k_1+k_2)t}}{\sqrt{v_1^{(J,n)}(T)v_2^{(J,m)}(T)}}
    \end{eqnarray}
    where $L_1=\sigma_1\int_0^te^{-k_1(t-s)}dW_1(s)+\sqrt{n}\nu_1e^{-k_1t}Z_1$ and $L_2=\sigma_2\int_0^te^{-k_2(t-s)}dW_2(s)+\sqrt{m}\nu_2e^{-k_2t}Z_2$.
\end{itemize}
\section{Calculation of  $p_{0,0}$, $p_{0,1}$, $p_{1,0}$ and $p_{1,1}$ }\label{app:integrals}
Based on the results in Cufaro Petroni and Sabino \cite{cs15} the joint \cdf\ of $X_1\sim\erl_1(\lambda_1)$ and $X_2\sim\erl_1(\lambda_2)$ is
\begin{equation}
    H(x_1,x_2) = \mathbbm{1}_{x_1\wedge\frac{x_2}{\gamma}\ge 0} \left[\left(1 - e^{-\lambda_1\left(x_1\wedge\frac{x_2}{\gamma}\right)}\right)-
    e^{-\lambda_2x_2}\left(1 - e^{-\left(\lambda_1-\gamma\lambda_2\right)\left(x_1\wedge\frac{x_2}\gamma\right)}\right)\right]
\end{equation}
For $\Delta t$ small we can assume that no more that one jump can occur hence:
\begin{eqnarray*}
  p_{1,1}  &=& \mathbb{P}(X_1\le \Delta t, X_2\le \Delta t) = H(\Delta t, \Delta t) \\
  p_{1,0} &=& \mathbb{P}(X_1\le \Delta t, X_2\ge \Delta t) = F_1(\Delta t) - H(\Delta t, \Delta t) \\
  p_{0,1} &=& \mathbb{P}(X_1\ge \Delta t, X_2\le \Delta t) = F_2(\Delta t) - H(\Delta t, \Delta t) \\
  p_{0,0} &=& \mathbb{P}(X_1\ge \Delta t, X_2\ge \Delta t) = 1 - p_{1,1} - p_{0,1} - p_{0,1}
\end{eqnarray*}
Finally, the results in section \ref{subsec:GOU} are obtained considering only $O(\Delta t)$ terms and splitting between $\gamma > 1$ and $ 0\le\gamma \le 1$.

\bibliographystyle{plain}
\bibliography{biblioCRM}

\end{document}